\documentclass[journal=nalefd,manuscript=article]{achemso}
\usepackage[version=3]{mhchem}
\input epsf.sty
\usepackage{amssymb,amsmath}
\usepackage{graphicx}
\usepackage[colorlinks]{hyperref}
\usepackage{achemso}

\newcommand{\captionfonts}{\small}

\makeatletter  
\long\def\@makecaption#1#2{%
  \vskip\abovecaptionskip
  \sbox\@tempboxa{{\captionfonts #1: #2}}%
  \ifdim \wd\@tempboxa >\hsize
    {\captionfonts #1: #2\par}
  \else
    \hbox to\hsize{\hfil\box\@tempboxa\hfil}%
  \fi
  \vskip\belowcaptionskip}
\makeatother   

\author{Samwel K. Sekwao}
\altaffiliation{Beckman Institute}
\author{Jean. P. Leburton}
\email{jleburto@illinois.edu}
\altaffiliation{Beckman Institute}
\affiliation[Department of Physics]
{Department of Physics}
\alsoaffiliation[Department of Electrical and Computer Engineering]
{Beckman Institute, University of Illinois at Urbana-Champaign, Urbana, Illinois 61801}

\title{Hot Electron Terahertz Oscillations in Graphene: \\Crater and Terraces in the Carrier Distribution Function.}

\abbreviations{IR,NMR,UV}
\date{\today}

\begin {document}

\begin{singlespace}
\begin{abstract}
In graphene, after the electric field is turned-on, the ballistic acceleration of charge carriers up to the monochromatic optic phonon energy generates a back-and-forth motion of the whole distribution function between the zero point energy and the phonon energy. This effect is predicted to manifest in damped terahertz oscillations of the carrier drift velocity and average energy. The most dramatic feature of this transient phenomenon is the onset of momentum-free areas surrounded by high momentum probability in phase space, and smooth steps or terraces in the distribution function. This dynamical effect that only takes place within a voltage and sample length window, is the direct consequence of the interplay between the electric force and the randomizing nature of deformation potential optic phonons in the linear band structure of graphene.
\end{abstract}

\leftline{Keywords: Hot carriers, graphene, Boltzmann transport, transient}
\newpage
Since its isolation in monolayer sheets \cite{Novoselov}, graphene has emerged as a new and ideal two-dimensional (2D) material with exotic electronic properties \cite{Katsnelson}. The particular nature of its band structure with a linear dispersion between energy and momentum $E = \hbar v_F|k|$, where $v_F \sim 10^8cm/s$ is the Fermi velocity of electron and holes, results in a zero-gap material, where the overlap between the valence and conduction band is reduced to a single point around the K and K' symmetry points of the Brillouin zone \cite{Katsnelson,Saito}. Hence, charge carrier dynamics can be described by a formalism similar to Dirac relativistic equation, where the speed of light is replaced by $v_F$ which provides new opportunity to observe ``quantum electro-dynamic'' effects in nanoscale systems. For electronics applications, graphene is similarly attractive because of the linear energy-momentum dispersion, for which all charge carriers move with the velocity, $v_F$ much larger than in conventional semiconductor materials \cite{Tiwari}, which anticipates faster time response.
\par
Indeed, the carrier mean free path for collisions with low energy acoustic phonons (AP) can reach several micrometers, while efficient scattering with monochromatic optic phonons (OP) occurs at much larger energy $(\hbar \omega_{op} \sim 0.2 eV )$ \cite{Piscanec} than in conventional semiconductors $(\hbar \omega_{op} \sim 0.04 eV )$ \cite{Kartheuser}.  Therefore, in electric fields $F$ high enough (to escape AP or low energy scattering), charge carriers can experience quasi-ballistic run-away until they scattered with efficient OP emission once $E \geq \hbar \omega_{op}$ \cite{Shockley}. The \textit{coherent} aspect of this process i.e. the coherent acceleration of the carrier distribution function followed by quasi-instantenous carrier relaxation by OPs at high energy is expected to produce oscillations of the carrier velocity with a periodicity given by $\tau_{gr} = \hbar \omega_{op}/eFv_F$ \cite{Matulionis}.
\par
In this paper we investigate carrier dynamics in graphene single layers, once the electric field has been turned-on, to determine the conditions of occurrence of current oscillations. Indeed, the thermal broadening of the initial distribution introduces a decoherence in the OP relaxation amongst carriers, which asides from low energy scattering, causes inherent damping of the oscillations, especially at high fields, where carriers overshoot the OP energy before emission. For this purpose, we use the Boltzmann formalism, and solve for the time-varying carrier distribution in the presence of OP scattering \cite{Matulionis,Brauer,J.P1}. We also account for low energy scattering such as impurities and acoustic phonons, by using the relaxation time approximation. In weak concentrations $(n_c \lesssim 10^{11}cm^{-2})$, electron-electron interactions do not play a major role in transport in graphene \cite{Castro}, and are not included in this analysis. We provide an analysis of the interplay between applied electric fields and strength of the low energy scattering rates for the onset of oscillations.
\par
\textbf{Model}. Let us consider a system of electrons in the graphene conduction band under the influence of an electric field along the x-direction. We divide the momentum space into two regions, I ($k < k_c$) and II ($k > k_c$) separated by a circle of critical momentum $k_c = \omega_{op}/v_F$, that corresponds to the electron kinetic energy equal to $\hbar \omega_{op}$ (Fig. 1a).

\begin{figure}[htp]
\centering
\vspace{0cm}
\epsfysize=5.5in
\epsfbox{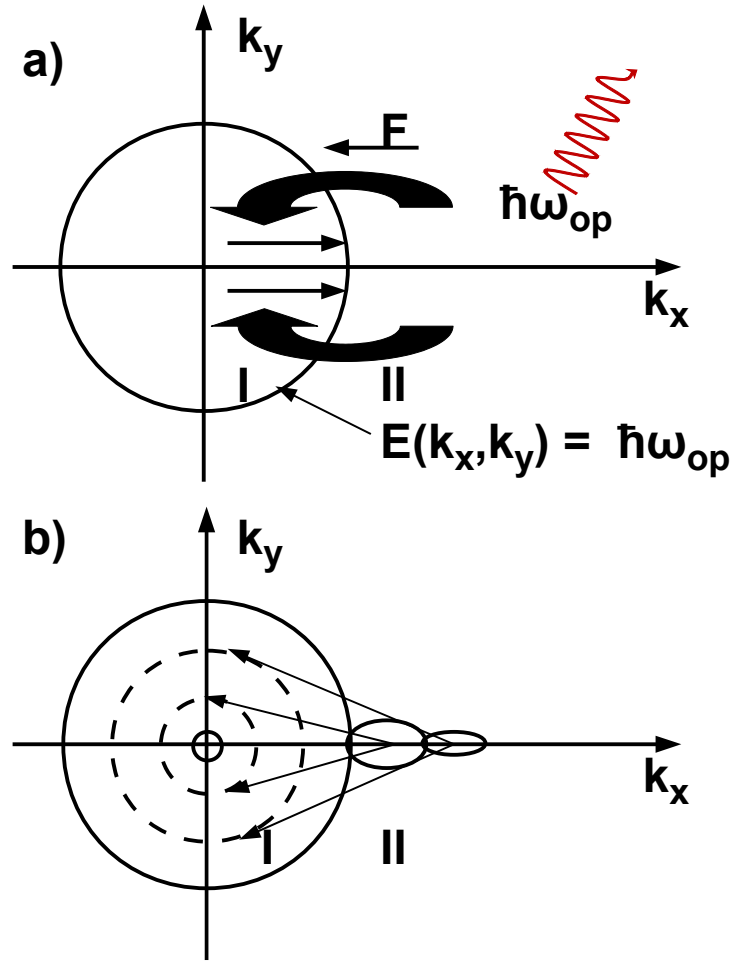}
\vspace{0cm}

\caption{ a) Schematics of carrier quasi-ballistic acceleration and OP scattering in 2D k space. The circle of radius $k_c$ is the locus of all points in $k$ space corresponding to the carrier energy $\hbar\omega_{op}$. b) Schematics of electrons scattered by OP's from two different positions of the distribution function in region II. The dotted circles in region I represent the two areas of high momentum probability where electrons are more likely to land.}

\end{figure}

\par
In region I electrons undergo quasi-ballistic acceleration and weak scattering by low energy mechanisms (e.g. impurities or APs) until they reach region II where they lose their energy by OP emission, and scatter back to region I. In our model, the electric field is assumed to be low enough so that electrons are scattered efficiently from region II to region I by OP emission with little probability to reach $E \geq 2\hbar \omega_{op}$.
\par
In the two regions, the time-dependent Boltzmann equations read:
\begin{equation*}\frac{\partial f_{I}(\vec k,t)}{\partial t} + \frac{eF} {\hbar} \frac{\partial f_{I}(\vec k,t)} {\partial k_x} = -\frac{f_{I}(\vec k,t) - f_o(\vec k)}{\tau} + \sum_{\vec k'} S(\vec k',\vec k)f_{II}(\vec k',t)\tag{1a}\end{equation*}
\begin{equation*}\frac{\partial f_{II}(\vec k,t)} {\partial t} + \frac{eF} {\hbar} \frac{\partial f_{II}(\vec k,t)} {\partial k_x} = -\frac{f_{II}(\vec k,t) - f_o(\vec k)}{\tau} - f_{II}(\vec k,t)\sum_{\vec k'} S(\vec k,\vec k')\tag{1b}\end{equation*}
where $f_{I}(\vec k,t)$ and $f_{II}(\vec k,t)$ are the time-dependent momentum distribution functions in regions I and II respectively, and $F$ is the electric field applied \cite{Leburton}. The first terms in RHS of equations (1) account for low energy scattering mechanisms where,
\setcounter{equation}{1}
\begin{equation}f_o(\vec k) = \frac{1}{1 + \exp(\frac{\hbar v_f k}{k_bT})}\end{equation}
is the Fermi-Dirac equilibrium distribution function, and $\tau$ is the relaxation time (for the sake of simplicity, we assume $\tau$ is $k$-independent and we vary it's value compared to the OP scattering rate).
$S(\vec k',\vec k)$ is the OP transition rate from a state with momentum $\vec k'$ to the state with momentum $\vec k$, given by \cite{Datta}
\begin{equation}S(\vec k,\vec k') = \frac{\pi{D_o}^2(N_q + 1)\delta(E' - E + \hbar \omega_{OP})}{
\sigma A \omega_{OP}}\end{equation}
where $D_o$ is the optical deformation potential, $\sigma$ is the mass of the graphene sheet per unit area, and $A$ is the area of the sheet.
\par
The second term on the right-hand side (RHS) of Eq. 1b is the carrier depopulation by OP emission, while in Eq. 1a it is the corresponding carrier repopulation at low energy. Here we neglect OP absorption processes since $\hbar\omega_{op}\gg k_BT$, even at room temperature where the phonon occupation number $N_q$ is negligible, so OPs only scatter electrons from region II to region I.
\par
\bigskip
We set the initial distribution $(t = 0)$ as the Fermi-Dirac distribution,
\begin{equation}f_I(k_x,k_y,t = 0) = f_o(k_x,k_y)\end{equation}
and solve Eqs.1 iteratively by noticing that for times $ 0 \leq t < \hbar k_c/eF$, the inside distribution drifts towards the critical circle with ``speed'' $eF/\hbar$. During this trip, one assumes that all carriers are in region I, corresponding to $f_{II} \approx 0$. Eq. 1a is just a differential equation for which a solution is readily obtained. Then we solve Eq. 1b for $k \geq k_c$, which is of the same form as Eq. 1a but with the additional OP emission term. Here we use the boundary condition $f_{I}(k = k_c, t) = f_{II}(k = k_c, t)$ \cite{Devreese}. For $t \geq \hbar k_c/eF$, we substitute the solution $f_{II}(\vec k,t)$ into the integral of the right hand side of Eq. 1a to start the same procedure for later times.
\par
\bigskip
As the electron population moves back and forth between regions I and II, each time undergoing more dephasing and broadening due to the finite duration of the OP emission process, we write the distribution function in each region at any time $t$ corresponding to the $n^{th}$  trip toward the critical circle, i.e. for $n= integer[eFt/\hbar k_c]+1$, as a superposition of distributions $f_I^{(i)}(k_x,k_y,t)$ and $f_{II}^{(i)}(k_x,k_y,t)$ of individual $i^{th}$ ``trip" ,
\begin{equation*}f_I(k_x,k_y,t) = \sum_{i = 1}^{n} f_I^{(i)}(k_x,k_y,t)\tag{5a}\end{equation*}
and
\begin{equation*}f_{II}(k_x,k_y,t) = \sum_{i = 1}^{n} f_{II}^{(i)}(k_x,k_y,t)\tag{5b}\end{equation*}
\setcounter{equation}{5}
We note that the concentration $n_c$ in the conduction band is found to be $n_c \approx 8$ x $10^{10} cm^{-2}$ for $E_F = 0$ (at the Dirac point) used throughout this analysis. For Fermi levels at $E_F = k_BT$ and $E_F = 2k_BT$ above the Dirac point, the carrier concentrations would be $n_c \approx 1.8$ x $10^{11} cm^{-2}$, and $n_c \approx 3.4$ x $10^{11} cm^{-2}$, respectively.

\begin{figure*}
 \centering
 \vspace{0cm}
 \epsfysize=4.15in
 \epsfbox{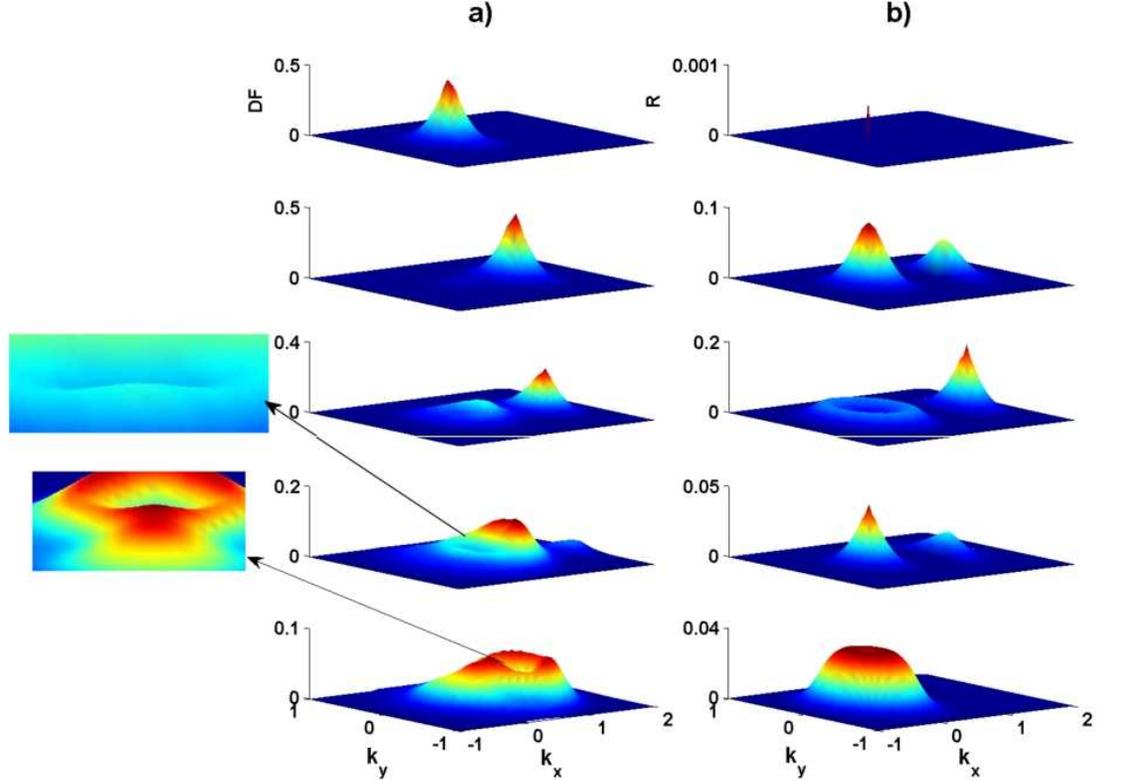}
 \vspace{0cm}
 \caption{ a) 3D plots of the distribution function at different times ($\beta$ = 0, 1, 1.5, 2, and 2.5) for $F = 1000V/cm$ and vanishing low energy scattering ($\gamma = 0$). b) Corresponding depopulation and repopulation rates for the same times as in (a)}
\end{figure*}

\par
\textbf{Results}. We introduce the dimensionless time parameter $\beta$ given by
\[ \beta = \frac{t}{t^*}\]
where
\[ t^* = \frac{\hbar k_c}{eF}\]
is the approximate time taken for the distribution to complete one ``trip''.
\par
Fig. 2a shows the time evolution of the distribution function distribution (DF) for $F = 1000 V/cm$ ($t^* \approx 2ps$), in the absence of low energy scattering ($\gamma = \tau_o/\tau = 0$). At $\beta = 0(t=0)$ the initial distribution is the tail of the Fermi distribution in the conduction band (Eq. 12), which is Maxwellian-like, and drifts along the $k_x$ direction (opposite to the $F$-field) to reach the OP energy at $\beta = 1$ (second  panel).  At that time, a ``hump'' appears around $k=0$ (not yet visible on the DF graph) as the front electrons reach region II, where they emit OP's and scatter back inside region I, as seen in the second panel of fig. 2b. As time progresses $\beta = 1.5$, the ``hump'' develops into a dimple, while the remaining high energy electrons from the first trip continue their drift in region II, where they experience strong OP depopulation. As seen in the third panel of fig. 2b, the corresponding repopulation rate at low energy exhibits a crater-like shape, which is due to the randomizing nature of the deformation potential OP scattering. Indeed, as schematically shown in fig.1b, after OP emission by high energy electrons, all $|\vec k- \vec k_c|$ -values become equiprobable, which forms a drifting circle of high repopulation rate, with increasing k-radius as first trip electrons penetrate deeper in region II.  This crater-like feature of the repopulation rate is the primary cause of the dimple in the low energy distribution function, which are areas in k-space where electrons have low probability to scatter. As time progresses, the DF ``dimple'' evolves into a crater-like shape, (fig.2.a, fourth and fifth panel), and the successive depopulation-repopulation OP processes overlapping at low energy with different amplitudes form also smooth terraces in the low energy tail of the distribution as it approaches steady state (fig.2.a , fourth panel). These morphological effects in electron distribution are unique to graphene as a result its linear band structure and the interplay of the quasi-ballistic acceleration and relaxation by high energy monochromatic OP's.

\begin{figure}[htp]
\centering
 \vspace{0cm}
 \epsfysize=4.0in
 \epsfbox{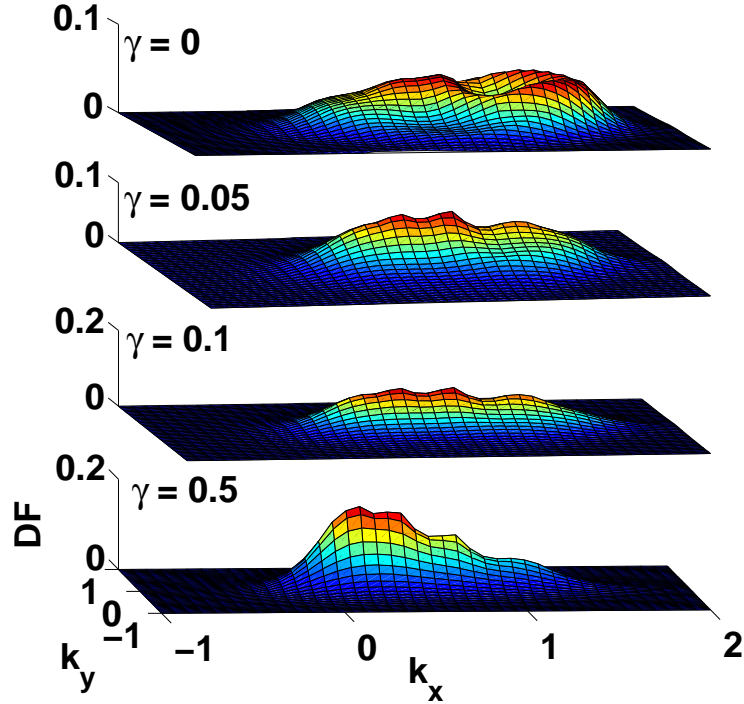}
 \vspace{0cm}

\caption{3D plots of the distribution functions at $\beta$ = 2.5 for $\gamma$ = 0, 0.05, 0.1, and 0.5. The applied field is $F = 1000V/cm$.}

\end{figure}
\par
Fig. 3 shows snapshots of the distribution function at $\beta = 2.5$ for varying low energy relaxation time, expressed in terms of $\gamma = \tau_0/\tau$. From the figure, one can see that the crater in the DF that occurs for $\gamma = 0$ progressively disappears as $\gamma$ increases. Indeed low energy scattering in region I re-distributes charge carrier momenta around $\vec k = 0$, especially in the crater center. For this reasons, the amplitude of the distribution in region I increases around $\vec k = 0$. Also, the distribution amplitude decreases in region II ($k > k_c$) as carriers spend more time in region I ($k \leq k_c$). Similarly the distribution recovers a streaming profile along the $k_x$-direction as conventional semiconductors \cite{Matulionis,Devreese}. Nevertheless, even for strong damping ($\gamma=0.5$), the distribution is characterized by a jagged profile, which contains the front and back ridges of the crater remnant still caused by the cumulative effects of the OP scattering for backward and forward carrier relaxation.
\begin{figure}[htp]
\centering
 \vspace{0cm}
 \epsfysize=4.0in
 \epsfbox{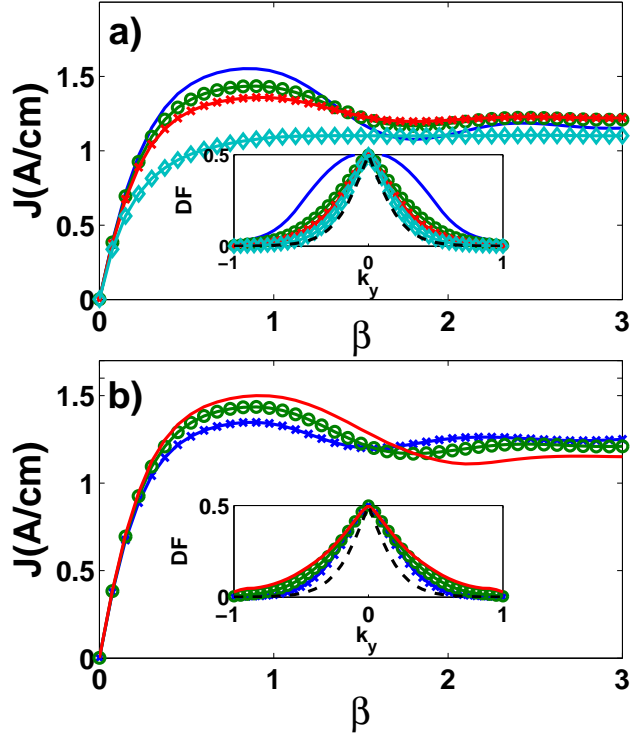}
 \vspace{0cm}

\vspace{0cm}

\caption{a) Current density as a function of time for different values of the low energy scattering parameters $\gamma$. Solid ($\gamma$  = 0), Circles ($\gamma$ = 0.05), Crosses ($\gamma$ = 0.1), and Diamonds ($\gamma$ = 0.5). The applied field is $F = 1000V/cm$. Inset: Cross-sections of the distribution at $\beta$ = 3 and $k_x = 0.5k_c$ for the corresponding values of $\gamma$. Arrows; initial distribution function at $k_x = 0$. b) Current density as a function of time for different fields and $\gamma$ = 0.1: Solid ($F = 1500V/cm$), Circles ($F = 1000V/cm$), Crosses ($F = 500V/cm$). Inset: Cross-section of the distribution function at $\beta$ = 3 and $k_x = 0.5k_c$ for the corresponding values of $F$. Diamonds; initial distribution function at $k_x = 0$.}

\label{fig1b}
\end{figure}

\par
The current density on the plane is given by,
\begin{equation} J_x(t) = -ev_F\sum_{\vec k}f(\vec k,t)\cos(\phi)\end{equation}
where $\phi$ is the angle between $\vec k$ and the $k_x$ axis.
Fig. 4a shows the current density as a function of time ($\beta$) for $F=1000 V/cm$ and different low energy scattering rates. For weak scattering, the current density overshoots its steady state value through damped oscillations, as a result of the back and forth motion of the distribution function (fig.1). The weaker the scattering, the higher the current overshoot. For strong low energy scattering, the current converges monotonically toward its steady state value without oscillations \cite{Matulionis}. Notice the stronger the scattering, the lower the steady state current value.  The insert shows the corresponding DF cross-sections at $k_x = 0.5$ and  $\beta = 3$ relative to the initial distribution at $k_x = 0$. As low energy scattering increases, the DF becomes narrower, taking a ``streaming'' profile in the electric field direction. This is due to the fact that as the low energy scattering increases, fewer electrons reach the high energy ( $E \geq \hbar \omega_{op}$) region II, reducing the number of electrons scattered back to the low energy region I, which narrows the distribution.
\par
Fig. 4b shows the current density as a function of time for three different field values, with $\gamma = 0.05$ in each case. One observes that the oscillation period $\tau_{gr} = \hbar\omega_{op}/eFv_F$ scales with the inverse of the electric field $F$. Also quite expectedly, the overshoot value increases with electric fields, but the damping is also enhanced with electric fields, which is due to the fact that the electron distribution penetrates the high energy($E \geq \hbar \omega_{op}$) region(II) farther than $\hbar \omega_{op}$, stretching the carrier relaxation by OP emission, which in turn broadens the DF in region I along the field, thereby reaching steady state quicker. The insert shows the DF cross-sections at $k_x = 0.5$ and $\beta = 3$ for the different fields, relative to the initial distribution at $k_x = 0$. As the field is increased, the distribution function broadens, because more electrons reach the high-energy ($E \geq \hbar \omega_{op}$ ) region II, and scatter back into region I broadening the distribution in the process.

\begin{figure}[htp]
\centering
 \vspace{0cm}
 \epsfysize=4.0in
 \epsfbox{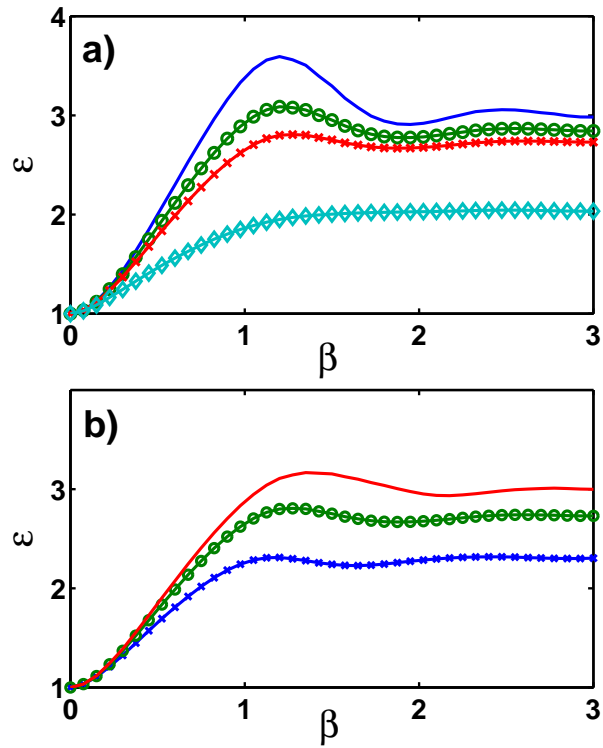}
 \vspace{0cm}

\caption{a) Normalized value of the carrier energy as a function of time for different values of the low energy scattering parameters $\gamma$. Solid ($\gamma$  = 0), Circles ($\gamma$ = 0.05), Crosses ($\gamma$ = 0.1), and Diamonds ($\gamma$ = 0.5). The applied field is $F = 1000V/cm$. b) Same but for different applied fields and $\gamma$ = 0.1: Solid ($F = 1500V/cm$), Circles ($F = 1000V/cm$), Crosses ($F = 500V/cm$).}

\label{fig1b}
\end{figure}
\par
The average energy density of carriers in the conduction band is given by,
\begin{equation} E(t) = \hbar v_F\sum_{\vec k}kf(\vec k,t)\end{equation}
where $f(\vec k,t)$ is the overall normalized distribution function, and the summation is taken over all $\vec k$ in regions I and II. At $\beta = 0$, the average energy per charge carrier is found to be $E(0) \approx 2.2k_B T$. This value is roughly twice larger than $k_BT$ expected for two-dimensional systems, and is the direct consequence of the linear energy dispersion in graphene, by contrast to the parabolic dispersion in normal 2D systems. Fig. 5a displays the ratio $\epsilon = E(\beta)/E(0)$ for $F = 1kV/cm$ and different low energy scattering rate ($\gamma$). As expected, the energy converges to higher values as low energy scattering is reduced. In addition, the convergence toward steady state occurs through damped oscillations, even for  $\gamma = 0$ as a consequence of the back and forth motion of the DF between the OP energy and the carrier zero-point energy. Quite clearly the oscillation period is given by $t^*$ for all $\gamma$. It is also seen that the oscillations persist even with significant low energy scattering. Fig. 5b displays the normalized energy $\epsilon$ as a function of time  for different fields and  $\gamma = 0.1$. As expected, carrier energies reach higher values as the field is increased, and the oscillation period decreases (larger $\beta$-period).
\par
\textbf{Discussions}. We have provided a transient analysis of the onset of current oscillations at the electric field turn-on caused by the back and forth motion of carrier distribution function between the zero-point energy and OPs in the presence of varying damping mechanisms. In this context we point out the anomalous shape of the carrier distribution as an interplay between ballistic acceleration and deformation potential OP emission in the transient regime. If OP-limited current oscillations have been predicted in GaAs, \cite{Matulionis} and indirectly observed in slightly n-doped InSb \cite{Lochner}, their manifestation in graphene is different in several respects: First, owing to the linear carrier energy-momentum dispersion, the oscillation periodicity is given by $\tau_{gr} = \hbar\omega^{gr}_{op}/eFv_F$ in graphene, while in GaAs parabolic conduction band  it is expressed as $\tau_{GaAs} = \sqrt{2m^*\hbar\omega^{GaAs}_{op}}/eF$, which nevertheless yields similar values, since the small effective mass, and the OP phonon frequency in GaAs compensate for the large $v_F$ \cite{Matulionis,J.P1}. Second, in III-V semiconductors, the OP polar nature focuses the low-energy repopulation along the $k_x$ axis, which provides a ``streaming'' profile to the carrier distribution instead of a crater-like shape in this case. Finally, in compound semiconductors, the oscillation onset is restricted by two conflicting conditions: On the one hand, the low value of the OP energy ($\hbar\omega_{op} : 0.04eV$) is comparable to the thermal broadening of the carrier distribution at room temperature so that the back and forth motion of the distribution between the optic phonon and the zero point energy is immediately damped \cite{Matulionis,Brauer,J.P1}. On the other hand, at low temperature, ionized impurity scattering becomes dominant and produces strong damping which can only be reduced by lowering the dopant density, thereby lowering the carrier density, and weakening the oscillation amplitude. In this respect, the high conductance of graphene, and the high optic phonon energy provide the conditions for room temperature observation.
\par
Low energy scattering should however still be minimized. Usually, for experimental studies and device applications, graphene-layers rest on a dielectric substrate, or are confined between two dielectric slabs \cite{Meric,Oostinga}.  In this case the presence of high K-dielectrics sandwiching the strictly 2-D graphene-layer may be used to screen charged impurity that may reduce low energy elastic scattering \cite{Jena}. However, dielectrics also contain interface and remote static charges that may offset dielectric screening \cite{Ponomarenko}. Moreover, the interaction between the 2D carriers in graphene and low energy Remote Interface Phonon (RIP) arising from the proximity of the substrate \cite{Hess,J.P2} introduces new scattering sources \cite{Fratini,Fischetti}. Therefore in general, suspended graphene-layers avoiding the RIP influence may be preferable \cite{Bolotin}.
\par
In this context to be observable at room temperature, the velocity oscillations should also take place within a parameter window. On the one hand the process requires $eV > \hbar\omega_{op}$ where $V$ is the external bias, to reach the OP energy, and on the other hand, the field should be large enough for carriers to escape low energy scattering, but $\hbar\omega_{op}/eFv_F\tau_{op} \gg 1$ where $1/\tau_{op}$ is the OP scattering rate, so that charge carriers don't penetrate the high-energy region $E \geq \hbar\omega_{op}$, scatter immediately after they reach the OP energy, which maintains the coherence of the distribution function. These requirements impose a lower and upper bound on the electric field, i.e. $0.5 kV/cm < F \ll 5 kV/cm$ (in graphene), and a lower bound on the sample length $L > v_F\tau_{op}$ ($>1\mu$m for $\tau_{op} < 1ps$), but $L$ should be smaller than a few values of  $\lambda = \hbar\omega_{op}/eF$, so as to prevent oscillation damping.
\par
One important issue for the validity of this analysis is the effect of leakage current due to electrons in the valence band crossing to the conduction of carriers through the Dirac point. Indeed it has been shown that there is still a minimum conductance ($G \sim 4e^2/h$) between the two bands despite the singular nature of the Dirac point \cite{Fradkin}. This value was later measured to be $G \sim e^2/h$ \cite{Novoselov}. However, this effect becomes important only when the graphene layer width $w \geq 23\mu m$, for which the band gap $E_g \leq 0.18meV$ are vanishing. Moreover, thermal effects such as acoustic phonon absorption by carriers in the valence band, are forbidden by conservation of both, energy and momentum. As for OP absorption we have shown earlier their occupation number is also negligible over the time scale considered in this analysis.
\par

\begin{acknowledgement}
Samwel K. Sekwao, thanks the Physics Department at the University of Illinois for continued support throughout this research project.
\end{acknowledgement}

\bibliography{mybib}

\end{singlespace}
\end{document}


\centerline{\textbf{Supporting Information}}
\textbf{Solution of Eqs. 1a \& b}. With the initial distribution $(t = 0)$ given the Fermi-Dirac distribution,
\begin{equation*}f_I(k_x,k_y,t = 0) = f_o(k_x,k_y)\tag{S.1}\end{equation*}
the solution for the ``first trip'' ($ 0 \leq t < \hbar k_c/eF$) in region I ($k \leq k_c$) is,
\begin{align*}f_I^{(1)}(k_x,k_y,t) & =  \frac{\nu_o}{\tau}\exp(-\frac{\nu_okx}{\tau})\int_{kx - \frac{t}{\nu_o}}^{kx}\,dz\exp(\frac{\nu_oz}{\tau})f_o(z,k_y) \nonumber \\
                                   & +  f_o(k_x - \frac{t}{\nu_o},k_y)\exp(-\frac{t}{\tau})\tag{S.2}\end{align*}
with
\[\nu_o = \frac{\hbar}{eF}\]
where $1/\nu_o$ is the ``speed'' in $k$-space with which the distribution drifts in the field towards the critical circle.
\par
\bigskip
\textbf{High Energy Distribution Function} ($k \geq k_c$). Since $\hbar v_fk \gg k_BT$ for $k > k_c$ even at room temperature, the approximation $f_o(k) \approx 0$ will be used for $E \geq \hbar \omega_{op}$. By setting a new variable $ u = t - \hbar k_x/eF$, Boltzmann equation (1.b) in region II reads,
\begin{equation*}\frac{\partial g_{II}(\vec k;u)}{\partial k_x} = -\nu_og_{II}(\vec k;u)\left\{\frac{1}{\tau} + \frac{1}{\tau_{op}(k)}\right\}\tag{S.3}\end{equation*}
where
\[g_{II}(\vec k;u) = g_{II}(k_x,k_y;u) = f_{II}(k_x,k_y;u + \nu_ok_x)\]
and the total OP emmission rate in Eq. 1b ($k > k_c$) is,
\begin{equation*}\frac{1}{\tau_{op}(k)} = \sum_{\vec k'} S(\vec k,\vec k') = \frac{{D_o}^2(k - k_c)}{2\sigma \hbar \omega_{OP}v_f} = \frac{1}{\tau_o}\left\{\frac{k}{k_c} - 1\right\} \tag{S.4}\end{equation*}
If we use $\hbar \omega_{op} = 0.2eV$, $D_o = 14eV/\AA$, $\sigma = 7.61$ x $ 10^{-7}kg/m^2$, then $\tau_o \approx 0.32p$sec. The solution of Eq. $S.3$ is then,
\begin{equation*}g_{II}(k_x,k_y;u) = g_{II}(k_x^o,k_y;u)\exp(-\Gamma(k_x,k_y))\tag{S.5}\end{equation*}
where
\[\Gamma(k_x,k_y) = \nu_o\int_{k_x^o}^{k_x}\,dk_x\left\{\frac{1}{\tau} + \frac{1}{\tau_{op}(k)}\right\}\]
and
\[k_x^o = \sqrt{{k_c}^2 - {k_y}^2}\]
The solution $(S.5)$ is valid only in the region $-k_c < k_y < k_c$, which is the region of interest. Going back to the original variables, the general solution reads,
\begin{equation*}f_{II}(k_x,k_y,t) = f_{II}(k_x,k_y,t -\nu_o(k_x - k_x^o)\exp(-\Gamma(k_x,k_y))\tag{S.6}\end{equation*}
The above solution describes a distribution function that drifts with ``speed'' $eF/\hbar$ in momentum space in the direction of the electric field, while decaying exponentially by OP emission.
\par
\bigskip
\textbf{Low Energy Distribution Function} ($k \leq k_c$). By using the same substitution as in the previous paragraph, the Boltzmann equation (1.a) inside the critical circle (region I) reads,
\begin{equation*}\frac{\partial g_I(\vec k;u)}{\partial k_x} = -\frac{\nu_o\{g_I(\vec k;u) - f_o(\vec k)\}}{\tau} + Z(\vec k;u)\tag{S.7}\end{equation*}
where
\[Z(\vec k;u) = \frac{{D_o}^2(k + k_c)}{4\pi \sigma \omega eFv_F}\int_{-\alpha(k)}^{\alpha(k)}\,d\phi'f_{II}(k + k_c,\phi',u + \nu_ok_x)\]
and
\[\alpha (k) = \arcsin \{\frac{k_c}{k + k_c}\}\]
The solution to $(S.7)$ is,
\begin{align*}g_I(\vec k;u) & =  \exp(-\frac{\nu_ok_x}{\tau})\int_{-k_x^o}^{kx}\,dz\exp(\frac{\nu_oz}{\tau})\left\{\frac{\nu_of_o(z,k_y)}{\tau} + Z(z,ky;u)\right\} \nonumber \\
                               & + \Omega(k_y;u)\exp(-\frac{\nu_ok_x}{\tau})\tag{S.8}\end{align*}
in which the function $\Omega(k_y;u)$ is determined by initial conditions. Going back to the initial variables,
\begin{align*}f_I(k_x,k_y,t) & =  \exp(-\frac{\nu_ok_x}{\tau})\int_{-k_x^o}^{kx}\,dz\exp(\frac{\nu_oz}{\tau})\left\{\frac{\nu_of_o(z,ky)}{\tau} + Z(z,k_y,t - \nu_ok_x)\right\} \nonumber \\
                               & + \Omega(ky,t - \nu_ok_x)\exp(-\frac{\nu_ok_x}{\tau})\tag{S.9}\end{align*}
By using the matching condition,
\begin{equation*}f_I(k_x^o,k_y,t) = f_{II}(k_x^o,k_y,t)\tag{S.10}\end{equation*}
and initial conditions, the solution for the distribution function in regions I and II is readily obtained.
\par
\bigskip
\textbf{Total Distribution Function}. Finally, we write the distribution function in each region at any time $t$ corresponding to the $n^{th}$  trip toward the critical circle, i.e. for $n= integer[eFt/\hbar k_c]+1$, as a superposition of distributions $f_I^{(i)}(k_x,k_y,t)$ and $f_{II}^{(i)}(k_x,k_y,t)$ of individual $i^{th}$ ``trip" ,
\begin{equation*}f_I(k_x,k_y,t) = \sum_{i = 1}^{n} f_I^{(i)}(k_x,k_y,t)\tag{S.11a}\end{equation*}
and
\begin{equation*}f_{II}(k_x,k_y,t) = \sum_{i = 1}^{n} f_{II}^{(i)}(k_x,k_y,t)\tag{S.11b}\end{equation*}
The normalization condition
\begin{equation*}n_c = \chi(t)\left \{\sum_{\vec k'}f_I(\vec k',t) + \sum_{\vec k'}f_{II}(\vec k',t)\right \}\tag{S.12}\end{equation*}
determines $\chi(t)$  as Eqs. ($S.11$) do not exactly conserve the number of carriers $n_c$. Indeed there is a decrease in the number of carriers in the system because at high fields electrons run away beyond region II i.e $E \geq 2\hbar \omega_{op}$ without being scattered. Another reason is that the summation in Eqs. ($S.11$) is limited to the $n = 3$ terms thereby ignoring the contribution of later ``trips''($n \geq 4$ terms) which might be relevant at higher fields considered in the analysis.